\documentclass[twocolumn,prl,amsmath,amssymb,showkeys,showpacs]{revtex4}

\usepackage{dcolumn}
\usepackage{graphicx}
\usepackage{bm}

\begin{document}

\title{Quantum Vacuum influence on the evolution of Pulsars}

\author{Arnaud Dupays}
\affiliation{Universit\'e de Toulouse, UPS, Laboratoire Collisions Agr\'egats R\'eactivit\'e,
           IRSAMC, F-31062 Toulouse, France \\
           CNRS, UMR 5589, F-31062 Toulouse, France}

\author{Carlo Rizzo} 
\affiliation{Laboratoire National des Champs Magn\'etiques Intenses (UPR 3228, CNRS-INSA-UJF-UPS),
           F-31400 Toulouse Cedex, France}

\author{Giovanni Fabrizio Bignami}
\affiliation{IUSS, Istituto Universitario di Studi Superiori, Pavia, Italy, EU}

\date{\today}

\begin{abstract}
In this letter we show that Quantum Vaccum Friction (QVF) 
should play an important role in neutron star evolution. 
Taking into account this effect we show that magnetars could be 
understood as a natural evolution of standard pulsars. 
For the Crab pulsar, of which the characteristic age is known, we present the
first completely coherent time evolution for its period and braking index.
For this pulsar we also give the predicted value of the current first derivative of
the braking index, providing a very important test to confirm QVF.
\end{abstract}

\maketitle

One of the most fundamental Quantum ElectroDynamics (QED) prediction concerns 
electric and magnetic properties of vacuum \cite{Euler,Rikken}. 
Following this idea, vacuum can be regarded as a standard medium 
with its own electromagnetic properties. Nevertheless, effects related to these properties 
are extremely weak and can only be tested in the presence of a very high electromagnetic field.
For that reason, pulsars are very appropriate systems to look for evidences for vacuum properties 
\cite{PRL.94,PRL.95}. 
Pulsars are fast-rotating neutrons stars, with a very high magnetic
dipole moment tilted with respect to their rotational
axis. Typically, their mass is of the order of the solar
mass, and their radius of the order of 10 km. But the most interesting feature of these 
stars is certainly their magnetic field which can exceed $10^8$ T on their surface \cite{ReviewPulsar}.
Pulsars are thus suitable laboratories to study magnetic properties of the surrounding vacuum.
In particular, a class of pulsars called magnetars \cite{Kouveliotou:2003} 
is characterized by a long spinning period and a 
magnetic field on the surface which is supposed to be even greater 
than the QED critical field $4\times10^9$ T \cite{Ibrahim,Bignami}.
At such a field non linear properties of quantum vacuum cannot be neglected. 
In general, the surface magnetic field of pulsars is inferred by comparing the measured period derivative
with the one obtained using the energy loss rate given by the classical formula for a radiating rotating magnetic dipole.
We believe that inferring a field exceeding the QED critical one using a pure classical formula 
(which neglects any QED effect) is questionable.
Heyl et al. \cite{Heyl.1,Heyl.2} have shown that the energy loss rate of a radiating rotating magnetic dipole, 
calculated by taking into account QED differs very little by the classical one. The use of a QED formula for the 
radiating dipole does not change in practise the inferred magnetic field on the surface of the pulsar.

Very recently, we have shown that a new energy loss process for pulsars should be taken into account, 
namely Quantum Vacuum Friction (QVF) resulting from the interaction between the 
non-stationary magnetic dipole moment of the star and its induced quantum vacuum 
magnetic dipole moment \cite{QVF}. In the case of magnetars, with very high magnetic fields, we have shown 
that the QVF is the dominating energy loss process and has important consequences. In particular, 
the inferred value of the magnetic field no long exceeds the QED critical field for all known magnetars. 

\begin{figure}[h]
\includegraphics[width=8cm,clip]{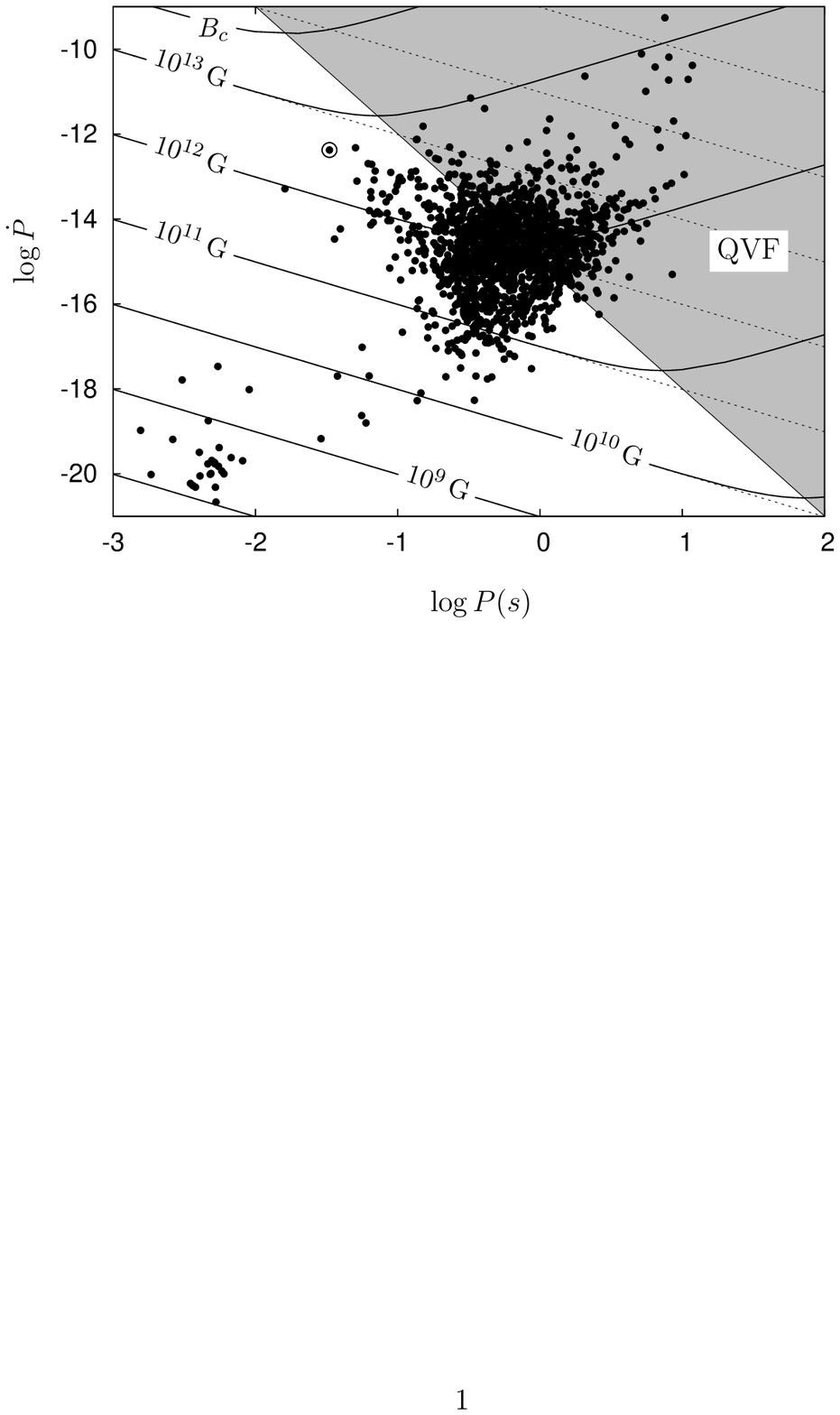}
\caption{\label{fig.1}{$P-\dot{P}$-diagram for the currently known pulsars. Pulsars located in the upper right 
area (gray) have an energy loss process dominated by QVF. The Crab pulsar is marked as a circle filled with a dot.}}
\end{figure}

\begin{figure}[h]
\includegraphics[width=8cm,clip]{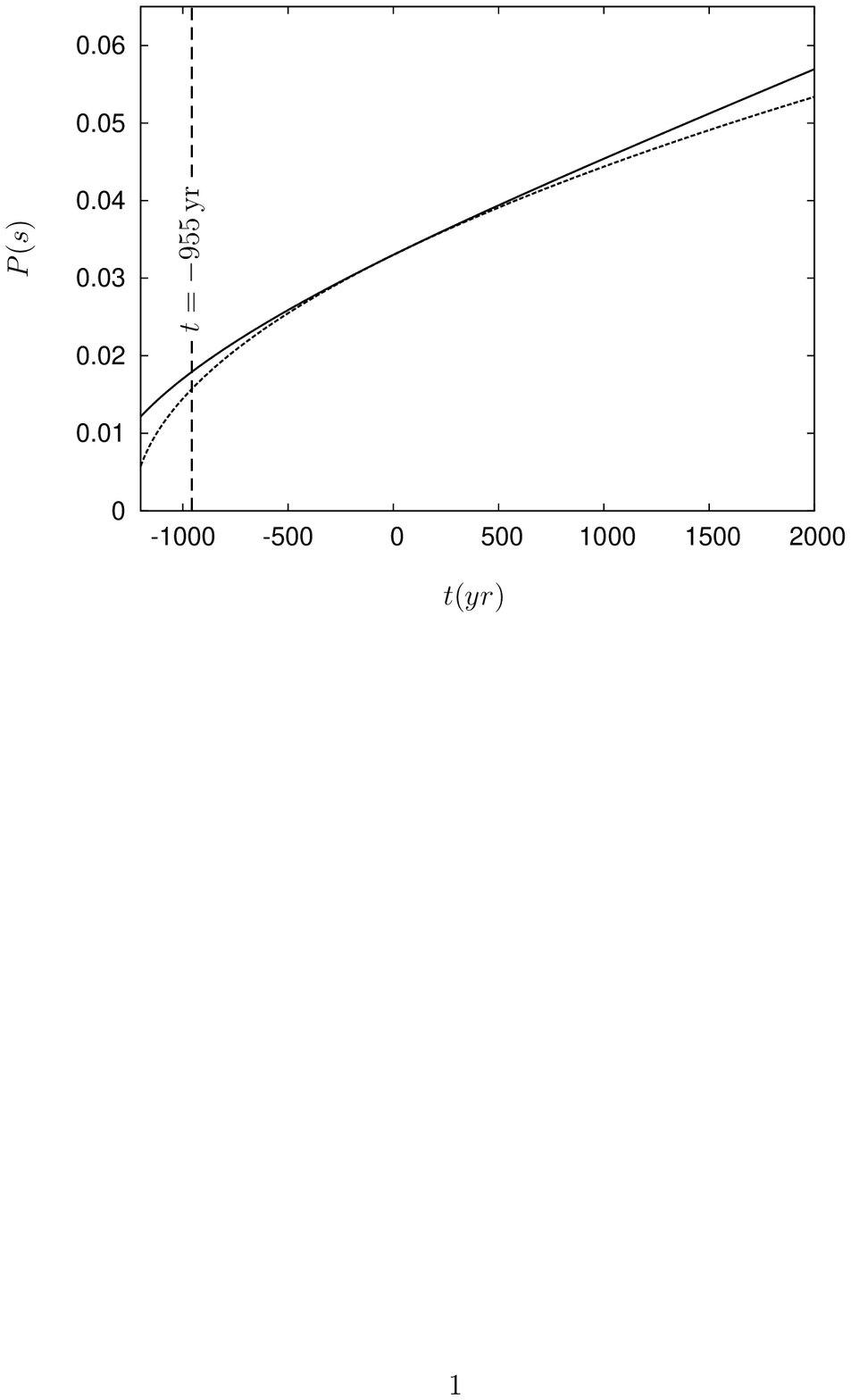}
\caption{\label{fig.2}{Spin period of the Crab pulsar as a function of time 
when QVF is taken into account (full line) and when QVF is neglected (dotted line). 
$t=0$ corresponds to current time (i.e "today").}}
\end{figure}

In this letter we consider some astrophysical consequences of 
QVF. We show that QVF plays a crucial role in neutron star evolution. Taking into account this 
effect, we show that magnetars can simply be understood as a natural evolution state of typical pulsars, 
similar to the Crab pulsar. For the Crab pulsar, of which the characteristic age is known, we present the 
first completely coherent time evolution for its period and braking index. 
Let us first summarize the main results concerning QVF in highly magnetized neutron stars. Considering 
a neutron star rotating in vacuum, we denote by $R$ its radius and $B_0$ the magnitude of the 
magnetic field at its surface. If $P$ is the spinning period of the neutron star and $\theta$ the inclination 
angle of its magnetic dipole moment with respect to the rotation axis of the star, one can show 
that for $B_0 \leq10^{10}$ T, the QVF energy loss rate $\dot{E}_{qv}$ is given by \cite{QVF}
\begin{equation}
  \label{eq.1}
  \dot{E}_{qv}\simeq
  \alpha\left(\frac{18\pi^2}{45}\right)\frac{\sin^2{\theta}}{B_c^2\mu_0c}\frac{B_0^4R^4}{P^2}
\end{equation}
where $B_c=m_e^2c^3/e\hbar\simeq 4.4\,10^{9}$ T is the QED critical field, $\alpha$ the fine 
structure constant, $\mu_0$ the vacuum permeability and $c$ the light velocity in vacuo. 
Total energy loss rate of the star $\dot{E}$ can thus be obtained by adding the classical dipolar 
radiative energy loss rate $\dot{E}_r$ given by 
\begin{equation}
  \label{eq.2}
  \dot{E}_{r}= \left(\frac{128\pi^5}{3}\right)\frac{\sin^2{\theta}}{\mu_0c^3}\frac{B_0^2R^6}{P^4}
\end{equation}
It is important to stress that the spinning period dependence is not
the same for the classical radiative process and for the QVF. 
The ratio of classical to QVF losses decreases like $1/P^2$ for large $P$, so 
that QVF becomes more important for slowly rotating neutron stars. 
Let us now study how QVF drastically changes the evolution of a pulsar. The evolution of 
a pulsar in period $P$ and slowdown $\dot{P}$ is usually described in a logarithmic $P-\dot{P}$-diagram 
as shown in fig.~\ref{fig.1}, where we have plotted all of the currently known pulsars ($\sim\,1700$) for 
which $P$ and $\dot{P}$ have been measured \cite{Pulsars}. On this diagram, we have also indicated lines of constant 
magnetic field in the case where only the classical radiative energy loss process is considered (dotted lines) 
together with the case for which QVF is taken into account (solid curves). These lines are obtained 
by matching the loss in rotational energy $\dot{E}$ with $\dot{E}_r+\dot{E}_{qv}$. Since $\dot{E}$ is given by 
\begin{equation}
\label{eq.3}
\dot{E}=4\pi^2I\dot{P}P^{-3}
\end{equation}      
where a neutron star moment of inertia of $I=10^{45}\,$g.cm$^2$ is assumed, we get
\begin{equation}
\label{eq.4}
\dot{P}=\left(\frac{32\pi^3}{3I}\right)\frac{B_0^2R^6}{\mu_0c^3P}+\alpha\left(\frac{3}{16I}\right)
\frac{B_0^4R^4P}{B_c^2\mu_0c}
\end{equation}  
where $\sin{\theta}$ is taken, as usual, equal to 1. 
The first term in eq.~(\ref{eq.4}) corresponds to the classical contribution, 
which gives the asymptotic behaviour for small $P$. The second term is the correction due to QVF, which becomes 
important for large $P$. In fig.~\ref{fig.1} we clearly distinguish two different evolution areas : pulsars located in 
the upper right area of the figure (gray) have an energy loss process dominated by QVF, 
whereas pulsars in the lower left area have classical magnetic dipole emission as the main energy loss. 
Defining the borderline between both areas as 
\begin{equation}
\label{eq.4a}
\frac{d\log{\dot{P}}}{d\log{P}}=0
\end{equation}
we get
\begin{equation}
\label{eq.4b}
\frac{\ddot{P}P}{\dot{P}^3}=0
\end{equation}
Since $P\neq0$, this equation implies that $\ddot{P}=0$ and thus,  
the borderline between both areas corresponds to pulsars having a constant slowdown $\dot{P}$. 
Assuming that the pulsar's magnetic field to remain constant during its evolution, we can describe the evolution of the 
star following the line on which it is situated today. Let us first ignore QVF. Since the rotational period of 
a pulsar increases with time, the natural evolution of these stars corresponds to a fall towards 
the lower right area of the figure along a given straight dotted line, depending 
of the value of their magnetic field. In this case, pulsars located in the upper right area 
can exhibit surface magnetic fields above the quantum critical field. Since these stars cannot be 
connected by a straight line of constant magnetic field to other pulsars with lower rotational period, we have to 
assume that these stars are very young, even if their rotational period is already high. For that reason, this class 
of pulsars, called magnetars, seems to be very different in nature from other standard pulsars. 
To understand the structure of these stars and the physical process involved which should give rise 
to such a huge surface magnetic field remains an open question. 
If we take into account QVF, we see that all magnetars are now connected by a modified line of constant 
magnetic field to pulsars with lower rotational period. Remarkably enough, their surface magnetic field becomes 
lower than the critical quantum field. We can then conclude, as far as QVF exists, that magnetars could be 
understood as a natural evolution of standard pulsars. Such a 
very important astrophysical consequence of quantum vacuum effect shows that testing the validity of the QVF theory 
would be crucial in the future to understand pulsars physics. It is also important to stress that with magnetic fields 
below the critical quantum field, the current explanation for the emmission properties 
of Anomalous X-ray Pulsars (AXPs) and Soft Gamma-ray Repeaters (SGRs) can not be applied to our model \cite{Qiao}, 
and so, this emmission remains an open question. 
Another interesting result of taking QVF into account in magnetized rotating neutron stars is a 
completely coherent time evolution of pulsars, as we will 
show in the following. Let us illustrate 
this point in detail with the Crab pulsar (the one marked as a circle filled with a dot in fig.\ref{fig.1}). 
Situated in the center of the Crab Nebula which is the remnant of the Super Novae SN1054, the Crab pulsar 
is certainly one of the most well-known pulsars \cite{Lyne.crab}. 
A very interesting feature of this pulsar is that historical records 
indicate its characteristic age to be $\sim 955$ yr. This indication, together with accurate
measurements of its current spin period $P$, first and second time derivative of
the spin period, respectively $P_1$ and $P_2$, drastically constrains its rotational 
evolution dynamics. Rewriting eq.~(\ref{eq.4}) as
\begin{equation}
\label{eq.5}
\dot{P}=\frac{K_r}{P}+K_{qv}P
\end{equation}
where $K_r=\dot{E}_r \times (P^4/4\pi^2I)$ and $K_{qv}=\dot{E}_{qv} \times (P^2/4\pi^2I)$, the age of a pulsar $\tau$ is 
given by 
\begin{equation}
\label{eq.6}
\tau=\int_{t_0}^{t}{dt^\prime}=\int_{P_i}^{P}{\frac{dP^\prime}{\frac{K_r}{P^\prime}+k_{qv}P^\prime}}
\end{equation} 
where $t$ denotes the current date and $P_i$ the initial spin period at $t=t_0$. We obtain 
\begin{equation}
\label{eq.7}
\tau=\frac{1}{2K_{qv}}\ln\left(\frac{K_r+K_{qv}P^2}{K_r+K_{qv}{P_i}^2}\right)
\end{equation}
If $K_{qv} \ll K_r$ eq.~(\ref{eq.7}) becomes
\begin{equation}
\label{eq.8}
\tau\simeq \frac{1}{2K_{r}}(P^2-{P_i}^2)
\end{equation}
From eq.~(\ref{eq.5}) we have in this case $K_r\simeq \dot{P}P$, so that
\begin{equation}
\label{eq.9}
\tau\simeq \frac{P}{2\dot{P}}\left(1-\left(\frac{P_i}{P}\right)^2\right)
\end{equation}
which is the expression usually employed to estimate the characteristic age of a pulsar. Inverting eq.~(\ref{eq.7}) 
(resp. eq.~(\ref{eq.9})) we obtain the spin period as a function of time when classical 
and QVF energy loss processes are taken into account (resp. when QVF is neglected). Both cases 
are presented in fig.~\ref{fig.2} for the Crab pulsar. In this figure, $t=0$ corresponds 
to current time (i.e "today"). Since the age of this pulsar is known, we can estimate the 
initial spin period $P_i$.   

Using the classical expression we obtain $P_i=16\,$ms, whereas 
we get $P_i=18\,$ms when QVF is taken into account. 
Since the energy loss process of the Crab pulsar is dominated by 
classical radiation (see fig.~\ref{fig.1}), we are not surprised 
that the global time evolution of this star does not drastically 
change when QVF effects are included. Nevertheless, the most 
important difference between both calculations is that the one 
taking into account QVF is the only one which provides a spin 
period evolution completely coherent with current observations, 
in particular with the current value of the braking index, 
as we will discuss in the following.

\begin{figure}[h]
\includegraphics[width=8cm,clip]{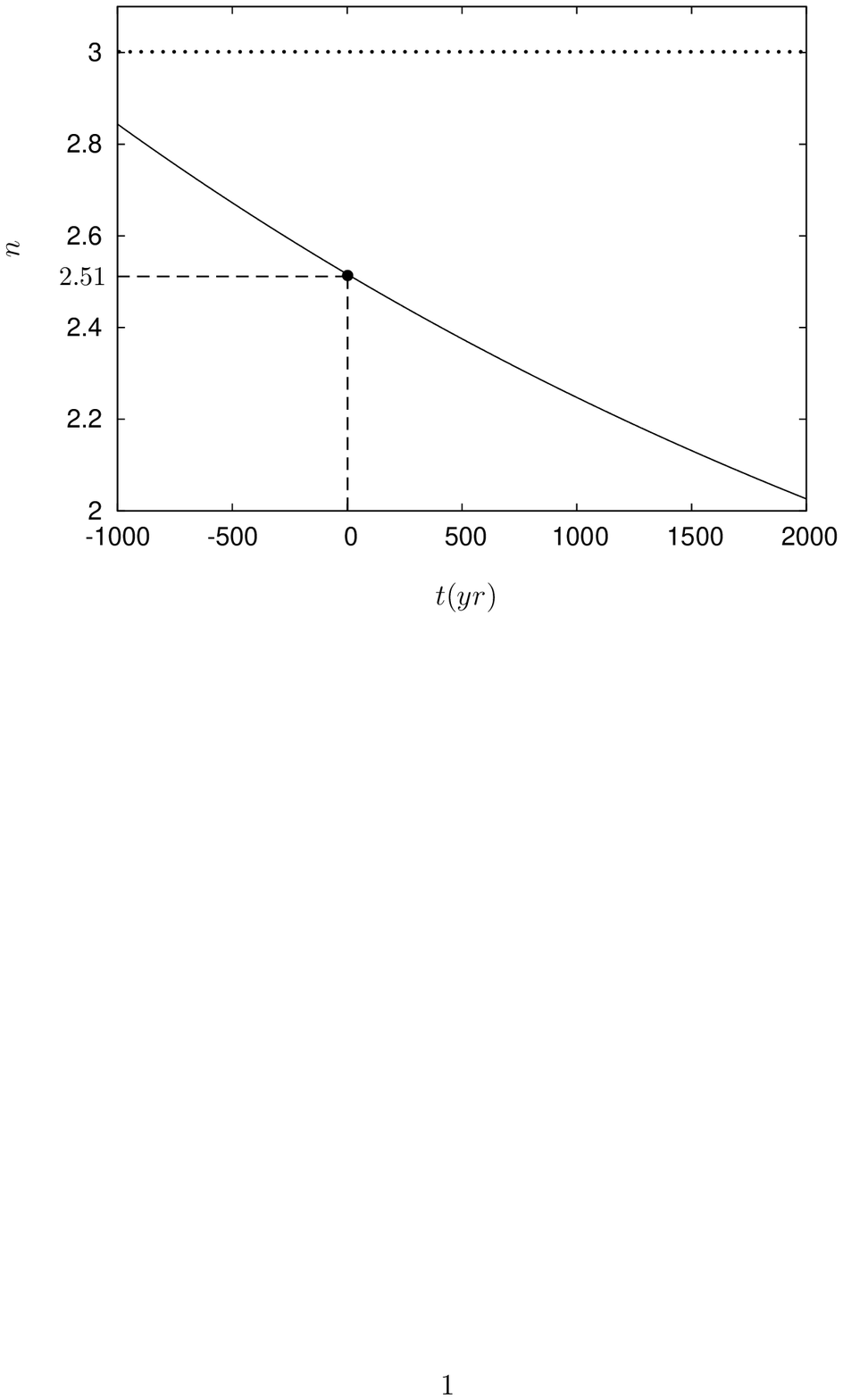}
\caption{\label{fig.3}{Braking index of the Crab pulsar as a function of time
when QVF is taken into account (full line) and when QVF is neglected (dotted line).
$t=0$ corresponds to current time.}}
\end{figure}
 
The braking index, $n$, 
is a fundamental dimensionless quantity 
given by \cite{Manchester}
\begin{equation}
\label{eq.10}
n=\frac{\nu\ddot{\nu}}{\dot{\nu}^2}
\end{equation} 
where $\nu$ is the spin frequency of 
the pulsar and $\dot{\nu}$ (resp. $\ddot{\nu}$), 
denotes the first (resp. the second) derivative 
of $\nu$ with respect to time. This parameter 
describes the rate at which a pulsar 
loses rotational energy. Introducing the spin 
period $P=1/\nu$, we rewrite the braking index as
\begin{equation}
\label{eq.11}
n=2-\frac{P\ddot{P}}{\dot{P}^2}
\end{equation}
From this expression we see that the borderline between gray and white plotted 
in fig.~\ref{fig.1} corresponds to pulsars having 
a braking index equal to 2 ($\ddot{P}=0$). 
Pulsars located in the upper right area of the figure (gray) 
have a braking index lower than 2,
whereas braking index of pulsars in the lower left area is higher.
In the following we study the time evolution 
of $n$ for $t>0$. 
We denote by $P_0$ the current spin period, and by 
$P_1$ and $P_2$, the current values of the first and the 
second derivative of $P$ with respect to time. The 
current value of $n$ is  
\begin{equation}
\label{eq.12}
n_0=2-\frac{P_0P_2}{P_1^2}
\end{equation}
At this stage it is important to stress that the constant value $n=3$ is 
expected for pure classical radiation models, whereas the current measured 
value is $n_0=2.51$ for the Crab pulsar. This discrepancy shows that pure 
classical radiation models would not lead to a realistic picture for this 
pulsar's time evolution. Adding QVF effects, solution of eq.~(\ref{eq.5}) 
becomes, for $t \geq 0$
\begin{equation}
\label{eq.13}
p^2(t)=\left(\frac{K_r}{K_{qv}}+P_0^2\right)e^{2K_{qv}t}-\frac{K_r}{K_{qv}}
\end{equation}
This expression allows us to calculate the first and the second derivative 
of $P$ with respect to time, and to match $\dot{P}(0)$ with $P_1$ and 
$\ddot{P}(0)$ with $P_2$, obtaining
\begin{equation}
\label{eq.14}
K_{qv}=\frac{P_0P_2+P_1^2}{2P_0P_1}
\end{equation}      
and 
\begin{equation}
\label{eq.15}
K_{r}=\frac{P_0P_1^2-P_2P_0^2}{2P_1}
\end{equation}
Finally, the time evolution of the spin period is given by
\begin{equation}
\label{eq.16}
P(t)=\sqrt{\frac{2P_0P_1}{A}\left(e^{At}-1\right)+P_0^2}
\end{equation}
with
\begin{equation}
\label{eq.17}
A=\left(\frac{P_2}{P_1}+\frac{P_1}{P_0}\right)
\end{equation}  
Using eq.~(\ref{eq.16}), we can calculate the braking index 
\begin{eqnarray}
\label{eq.18}
n(t)&=&1-\left(\frac{P_0P_2}{P_1^2}-1\right)e^{-At} \nonumber \\
&=&1-(1-n_0)e^{-At}
\end{eqnarray}
The braking index time evolution for the Crab pulsar is presented 
in fig.~(\ref{fig.3}). When QVF is taken into account, the braking index 
is a decreasing function with respect to time which is consistent with 
a global rotation slowdown. From eq.~(\ref{eq.18}) 
we can predict the current value of the first derivative of $n$ given 
by 
\begin{equation}
\label{eq.19}
\dot{n}(0)=A(1-n_0)
\end{equation}
As far as we know, there are only two pulsars in the ATNF catalogue \cite{Pulsars} 
for which the calculated value of $\dot{n}^c(0)$ can be compared with our estimation given 
by eq.~(\ref{eq.19}). These stars are the Crab pulsar and the B1509-58 pulsar. 
For the Crab pulsar we get $\dot{n}(0)\sim-1.8421\,10^{-11}\,$s$^{-1}$ and 
$\dot{n}^c(0)\sim1.5676\,10^{-13}\,$s$^{-1}$, whereas for the B1509-58 pulsar 
we get $\dot{n}(0)\sim-1.3550\,10^{-11}\,$s$^{-1}$, a value in perfect agreement with the 
calculated one $\dot{n}^c(0)\sim-1.2500\,10^{-11}\,$s$^{-1}$. These results show that new 
measurements of $\dot{n}(0)$ are needed to improve statistics and thus provide a very important 
test to confirm our QVF predictions. 

In this letter we have shown that QVF should play an important role 
in neutron star evolution. Taking into account this effect, we have shown 
that the surface magnetic field of magnetars becomes lower than the critical 
quantum field, so that magnetars could be simply understood as a natural evolution 
of standard pulsars. For pulsars for which the braking index is known, QVF provides 
a completely coherent spindown history, as illustrated with the Crab pulsar. 
For this pulsar we have also given the predicted value of the current first derivative of 
the braking index, suggesting a very important test to confirm QVF.

\end{document}